\documentclass[prl,twocolumn,superscriptaddress,showpacs]{revtex4}
\begin{document}
\author{I.M. Sokolov} 
\affiliation{Institut f\"ur Physik, Humboldt-Universit\"at zu Berlin,
Newtonstr. 15, D-12489, Berlin, Germany}
\author{M.G.W. Schmidt}
\affiliation{Institut f\"ur Physik, Humboldt-Universit\"at zu Berlin,
Newtonstr. 15, D-12489, Berlin, Germany}
\affiliation{Departament de Qu\'imica F\'isica, Universitat der Barcelona,
Mart\'i i Franqu\`es 1, E-08028, Barcelona, Spain}
\author{F. Sagu\'es} 
\affiliation{Departament de Qu\'imica F\'isica, Universitat der Barcelona,
Mart\'i i Franqu\`es 1, E-08028, Barcelona, Spain} 
\title{On reaction-subdiffusion equations} 

\date{\today}

\begin{abstract}
To analyze possible generalizations of reaction-diffusion schemes for
the case of subdiffusion we discuss a simple monomolecular conversion
$A \rightarrow B$. We derive the corresponding kinetic equations for
local $A$ and $B$ concentrations. Their form is rather unusual: The
parameters of reaction influence the diffusion term in the equation
for a component $A$, a consequence of the nonmarkovian nature of
subdiffusion.  The equation for a product contains a term which
depends on the concentration of $A$ at all previous times.  Our
discussion shows that reaction-subdiffusion equations may not
resemble the corresponding reaction-diffusion ones and are not obtained 
by a trivial change of the diffusion operator for a subdiffusion one.
\end{abstract}


\maketitle  

Many phenomena in situations out of equilibrium can be described using
a picture based on reaction processes. Apart from chemical reactions, 
the examples
are exciton quenching, recombination of charge carriers or radiation
defects in solids, predator-pray relationships in ecology
etc. Reactions in homogeneous media are often described by formal
kinetic schemes. Thus, the concentrations $C_i(t)$ of the components
 follow the first-order
differential equations $d C_i(t)/ dt = f_i\{C_1(t),...,C_N(t)\}$ where
the reaction terms typically have a form $f_i\{C_1,...,C_N\}= \pm
\kappa_i C_1^{n_1} C_2^{n_2} ... C_1^{n_N}$ with the powers $n_j$
depending on the stoichiometry of the reaction and $\kappa_i$
denoting the corresponding reaction rates.  In inhomogeneous situations
(layered systems, fronts, etc.)  the mesoscopic approach based on
reaction-diffusion equations for the position-dependent concentrations
$C_i(\mathbf{r},t)$ is often the appropriate way of description. In
case of normal diffusion such equations are obtained by adding a
diffusive term to classical reaction schemes and have the form
\begin{equation}
\frac{\partial C_i(\mathbf{r},t)}{\partial t} = K_i \Delta
C_i(\mathbf{r},t)+ f_i
\label{RDS}
\end{equation}
with $f_i=f_i\{C_1(\mathbf{r},t),...,C_N(\mathbf{r},t)\}$ and $K_i$
being the diffusivity of the component $i$.  This approach
is applicable whenever characteristic scales of spatial
inhomogeneities are much larger than the typical interparticle
distances and particles' mean free paths (see e.g.\cite{MyBook}). 
As we proceed to show, the possibility to put down such schemes is due 
to the Markovian nature of normal diffusion.

Nowadays more and more attention is paid to situations when the
diffusion is anomalous, which are found to be abundant
\cite{PhysWorld}.  One of the most important situations here is the
case of subdiffusion described within the continuous-time random walk
(CTRW) scheme \cite{MetzKlaf}.  In this case the subdiffusive nature
of motion stems from the fact that particles get trapped and have to
wait for a time $t$ (distributed according to power-law probability
density function $\psi(t) \propto t^{-1-\alpha}$) until the next step
can be performed.  It was shown that the properties of the reaction
under such subdiffusion might be vastly different from those in
diffusive systems \cite{Katjas,Katjas2,Silbeys,Seki}.  The microscopic
approach of these works aims on the understanding of the situation
when the particles performing CTRW react on encounter (and don't react
as long as they do not move). Such situation is pertinent to exciton
quenching in solids, or to transport in ion channels \cite{Goychuk}.

In many cases, however, a mesoscopic approach is desirable. Such an
approach was adopted in the case of reactions under superdiffusion due
to L\'evy flights, Refs.\cite{Chen,DelCast}, where the transport
process involved is Markovian. The situation with subdiffusion is much
more subtle due to strongly nonmarkovian character of subdiffusive
transport \cite{Shushin}.  Here two different situations can be
encountered: either the reaction at small scales is also
subdiffusion-controlled (like in the models discussed above, where
particles can only react if a new step is made) or it locally follows
normal, classical kinetics.  
This last case that we address here is physically relevant
since it describes reactions in porous media. The situation is of
extreme importance in hydrology, where the transport in catchments
is hindered by trapping in stagnant regions of the flow,
caves and pores on all scales.  The transport at long times and large
scales is adequately described by CTRW \cite{Schers}. However on
small scales reactions take place in normal aqueous solutions,
so that particles trapped in stagnant regions still can react with
each other.  A mesoscopic approach to such a case was adopted in
\cite{FeMe} within a probabilistic scheme, while \cite{Wearnes} tackle
this problem by using equations of the same form as our Eq.(\ref{RDS})
where the diffusion operator is changed for a subdiffusion one, containing 
an additional fractional derivative in time
\cite{MetzKlaf,Phys2day}:
\begin{equation}
\frac{\partial C_i(\mathbf{r},t)}{\partial t} = \:_0
D_t^{1-\alpha_i} K_{i,\alpha_i} \Delta C_i(\mathbf{r},t) + f_i \:.
\label{RDF}
\end{equation}
In this equation $\alpha_i$ is the exponent of the anomalous diffusion for the
component $i$, $\:_0 D_t^{1-\alpha_i}$ is the operator of fractional
(Riemann-Liouville) derivative, and $ K_{i,\alpha_i}$ is the
corresponding anomalous diffusion coefficient. Such equations with
decoupled transport and reaction term  were postulated based on the 
analogy with Eq.(\ref{RDS}) and look quite plausible. In some cases also 
a reaction term has to be modified by applying a 
fractional derivative as suggested by a microscopic model in \cite{Katjas2}.   

In what follows we derive the reaction-subdiffusion equations for the
simplest reaction scheme (monomolecular conversion $A \rightarrow B$)
corresponding e.g. to radioactive decay of isotope $A$ which is
introduced into the ground water at some place at time $t=0$ and is
transported according to anomalous diffusion.  We show that the
corresponding equations \textit{do not} follow a pattern of
Eq.(\ref{RDF}), so that the reaction and diffusion terms do not
decouple.

Let us assume for the time being that all properties of
$A$ and $B$ particles are the same, so that the reaction corresponds to a
relabeling of $A$ into $B$ taking place at a rate $\kappa$. 
In what follows we will use one-dimensional notation, the generalization 
to higher dimensions is trivial. The Eqs.(\ref{RDS}) for this case read:
\begin{equation}
\frac{\partial A}{\partial t}  = K\Delta A-\kappa A, \qquad
\frac{\partial B}{\partial t} = K\Delta B+\kappa A.
\label{E1E2}
\end{equation}
with $K$ being the normal diffusion constant.
Let $C(x,t)=A(x,t)+B(x,t)$ be the sum of concentrations. It evolves
according to a diffusion equation:
\begin{equation}
\frac{\partial C}{\partial t}=K\Delta C.
\end{equation}
Both concentrations $A$ and $B$ follow
\begin{equation}
A(x,t) = e^{-\kappa t}C(x,t), \; B(x,t) = \left[
1-e^{-\kappa t}\right] C(x,t).
 \label{E3}
\end{equation}
To see this apply Laplace-transform to the equations
(\ref{E1E2}). Note that the solution for $A(x,t)$ in the Laplace domain
is $\tilde{A}(x,u)=\tilde{C}(x,u+\kappa )$.  Eq.(\ref{E3}) reflects
the fact, that the conversion is independent from the motion of
particles, so that concentrations of $A$s and of $B$s are
proportional to the overall concentration multiplied by the
probability for a particle to survive as $A$ or to become $B$. The
same argument leads to the conclusion that Eq.(\ref{E3})
also holds in anomalous diffusion,
whatever the evolution equation for $C$ is. For subdiffusion
\begin{equation}
\frac{\partial C(x,t)}{\partial t}=K_{\alpha }\,_{0}D_{t}^{1-\alpha
}\Delta C(x,t)
\label{Sum}
\end{equation}
so that in Fourier-Laplace domain for the initial condition
$C(x,0)=\delta (x)$ one has 
$\tilde{C}(k,u)=\left(u+u^{1-\alpha }k^{2}K_{\alpha}\right)^{-1}$
so that, for instance
\begin{equation}
\tilde{A}(k,u)=\frac{1}{\left( u+\kappa \right) +\left( u+\kappa
\right) ^{1-\alpha }k^{2}K_{\alpha}}.  \label{Result}
\end{equation}
However, neither the solution of Eq.(\ref{RDF}) nor the solution of
the fractional equation with the modified reaction term \cite{Katjas2} 
reproduce this result: the simple
reaction-subdiffusion schemes do not describe the conversion
reaction correctly.  

Let us now turn to deriving a correct reaction-diffusion equations for
our case.  Our derivation will follow the way of derivation of the
generalized master equation for CTRW used in ref. \cite{CGS} based on
the ideas of \cite{Bur}.  We start from a discrete scheme and consider
particles occupying sites of a one-dimensional lattice.  The
generalized reaction (sub-)diffusion equations follows from the
balance conditions for particle numbers.  A balance condition for the
mean number $A_{i}$ of particles $A$ on site $i$ of the system reads
\begin{equation}
\frac{dA_{i}(t)}{dt}=I_{i}^{+}(t)-I_{i}^{-}(t)-\kappa A_{i}(t)
\label{sitebalance}
\end{equation}
where $I_{i}^{-}(t)$ is the loss per unit time due to the particles'
departure from the site (loss flux) at site $i$, $I_{i}^{+}(t)$ is the
gain flux, and $\kappa A_{i}$ is the loss due to
conversion. Particles' conservation for transitions between the two
neighboring sites corresponds to
\begin{equation}
I_{i}^{+}(t)=w_{i-1,i}I_{i-1}^{-}(t)+w_{i+1,i}I_{i+1}^{-}(t),
\label{Acontinuity}
\end{equation}
where $w_{i,j}$ is a probability to jump to site $j$ when leaving
$i$. For unbiased walks one has $w_{i-1,i}=w_{i+1,i}=1/2$. Thus:
\begin{equation}
\frac{dA_{i}(t)}{dt}=
\frac{1}{2}I_{i-1}^{-}(t)+\frac{1}{2}
I_{i+1}^{-}(t)-I_{i}^{-}(t)-\kappa A_{i}(t).  
\label{balance1}
\end{equation}
We now combine this continuity equation with the equation for
$I_{i}^{-}(t)$ following from the assumption about the distribution of
sojourn times.  The loss flux at time $t$ is connected to the gain
flux at the site in the past: the particles which leave the site $i$
at time $t$ either were at $i$ from the very beginning (and survived
without being converted into $B$), or arrived at $i$ at some later
time $t^{\prime }<t$ (and survived). A probability density that a
particle making a step at time $t$ arrived at its present position at
time $t^{\prime }$ is given by the waiting time distribution $\psi
(t-t^{\prime})$, the survival probability being $p(t)=e^{-\kappa t}$. Thus:
\begin{equation}
I_{i}^{-}(t) = \psi (t)e^{-\kappa t}A_{i}(0)+\int_{0}^{t}\psi
(t-t^{\prime })e^{-\kappa (t-t^{\prime })}I_{i}^{+}(t^{\prime
})dt^{\prime }.
\end{equation}
Applying Eq.(\ref{sitebalance}) we get:
\begin{eqnarray}
I_{i}^{-}(t) &=& \psi _{s}(t)A_{i}(0) \\
&+& \int_{0}^{t}\psi _{s}(t-t^{\prime })\left[ \frac{A_{i}(t^{\prime
})}{d t^{\prime}}+\kappa A_{i}(t^{\prime }) + I_{i}^{-}(t^{\prime })\right]
dt^{\prime },\nonumber
\label{main}
\end{eqnarray}
where $\psi _{s}(t)=\psi (t)e^{-\kappa t}$ is the non-proper waiting
time density for the actually made new step provided the particle
survived.  This approach can also be generalized to bimolecular
reactions \cite{generalbimol}.  Changing to the Laplace domain and
noting that $\tilde{\psi}_{s}(u)=\tilde{\psi}(u+\kappa )$ we get
\begin{equation}
\tilde{I}_{i}^{-}(u)= 
\tilde{\Phi}_{\kappa }(u)\tilde{A}_{i}(u)
\label{currentAlaplace}
\end{equation}
with $\tilde{\Phi}_{\kappa }(u)$ given by
\begin{equation}
\tilde{\Phi}_{\kappa }(u) = \frac{(u+\kappa )\,\tilde{\psi}(u+\kappa )}{1-
\tilde{\psi}(u+\kappa )}.
\end {equation}
Returning to the time-domain we thus get
\begin{equation}
I_{i}^{-}(t)=\int_{0}^{t}\Phi_{\kappa}(t-t^{\prime })A_{i}(t^{\prime
})dt^{\prime },
\label{IntEq}
\end{equation}
Note that $\Phi_{\kappa}(t)$ given by
the inverse Laplace transform of $\tilde{\Phi}_{\kappa }(u)$ corresponds
to $\Phi_{\kappa}(t)=\Phi_0(t)e^{-\kappa t}$ where 
$\Phi _{0}(t)$ obtained by taking $\kappa=0$ is the usual
memory kernel of the generalized master equation for CTRW.

Combining Eq.(\ref{IntEq}) with Eqs.(\ref{sitebalance}) and 
(\ref{Acontinuity}) we get:
\begin{eqnarray}
\frac{d A_{i}(t) }{dt}&=&\int_{0}^{t}\Phi _{\kappa
}(t-t^{\prime })\left[ \frac{A_{i-1}(t^{\prime
})}{2}+\frac{A_{i+1}(t^{\prime })}{2}\right.- \nonumber \\
&-& A_{i}(t^{\prime })\Big]dt^{\prime }-\kappa A_{i}(t).
\end{eqnarray}
Transition to a continuum in space ($x=ai$) gives
\begin{eqnarray}
\frac{\partial A(x,t) }{\partial t} = \frac{a^{2}}{2}\int_{0}^{t}\Phi
_{\kappa }(t-t^{\prime })\Delta A(x,t^{\prime })dt^{\prime }-\kappa
A(x,t)  \nonumber \\
= \frac{a^{2}}{2}\int_{0}^{t}\Phi
_{0}(t-t^{\prime })e^{-\kappa (t-t^{\prime })}\Delta A(x,t^{\prime
})dt^{\prime }-\kappa A(x,t),
\end{eqnarray}
a rather unexpected form, where the reaction rate explicitly affects
the transport term.  

For the exponential waiting time distribution
$\psi(t)=e^{-\lambda t}$ corresponding to $\tilde{\psi} (u)=\lambda
/ (u+\lambda )$ the kernel reads $\Phi_{0}(t) =\lambda \delta(t)$, and
the existence of an additional exponential multiplier does not play any role: 
The reaction diffusion equation is perfectly exact.  

In the case of slowly decaying $\Phi_{0}(t)$ the exponential cutoff
introduced by the reaction is crucial.  For power-law waiting time
distributions and for $\kappa =0$ the integral operator
$\int_{0}^{t}\Phi_0(t-t^{\prime})f(t^{\prime })dt^{\prime }$ is the
operator of the fractional derivative: For such distributions 
$\tilde{\psi}(u)\simeq 1-\left( \tau u\right) ^{\alpha}\Gamma(1-\alpha)$ 
(where $\tau $ is the appropriate time scale) and
(for $u\rightarrow 0$) we have $\tilde{\Phi}_{0}(u)\simeq
(1/\tau^{\alpha}\Gamma(1-\alpha))u^{1-\alpha }$ which is proportional to
the operator of the Riemann-Liouville derivative of the order $\alpha$: 
$\frac{a^{2}}{2}\int_{0}^{t}\Phi_{0}(t-t^{\prime})f(t^{\prime
})dt^{\prime } = \,K_{\alpha}\,_{0}D_{t}^{1-\alpha }f$ for
sufficiently regular functions $f$. The generalized diffusion coefficient reads
$K_{\alpha}=a^2[2\tau^{\alpha}\Gamma(1-\alpha)]^{-1}$.  
For $\kappa>0$ however the reaction affects the diffusion part of the equation:
the Laplace transform of the integral kernel $\Phi_{\kappa}(t)$ reads 
\begin{equation}
\tilde{\Phi}_{\kappa}(u) \simeq \frac{1}{\tau^{\alpha}\Gamma(1-\alpha)}
(u+\kappa )^{1-\alpha }
\end{equation}
and is no more a fractional derivative.
The integral operator 
$\hat{T}_{t}(1-\alpha,\kappa)f= \tau^{\alpha}\Gamma(1-\alpha)
\int_{0}^{t}\Phi_{\kappa }(t-t^{\prime }) f(t^{\prime })dt^{\prime }$
corresponds in time domain to
\begin{eqnarray}
\hat{T}_{t}(1-\alpha,\kappa)f&=&
\left(\frac{d}{dt}
\int_{0}^{t}\frac{e^{-\kappa(t-t^{\prime })}}{(t-t^{\prime })^{1-\alpha }}
f(t^{\prime})dt^{\prime }\right. \nonumber \\
&+& \left. \kappa \int_{0}^{t}\frac{e^{-\kappa (t-t^{\prime
})}}{(t-t^{\prime })^{1-\alpha }}f(t^{\prime })dt^{\prime }\right),
\label{operatorT}
\end{eqnarray}
turning to be a fractional derivative only for $\kappa=0$.
 The equation for the $A$-concentration thus finally reads:
\begin{equation}
\frac{\partial A(x,t) }{\partial t}= K_{\alpha} \hat{T}_{t}(1-\alpha,\kappa)
\Delta A(x,t)-\kappa A(x,t).
\label{Aconc}
\end{equation}
Although our reaction does not depend on the particles' motion, the parameters
of the reaction explicitly enter the transport operator $\hat{T}$ of the 
equation.

Analogously we will now derive an equation for the $B$-particles. As
for $A$ one has a balance condition for the mean particle number
$B_{i}$ of $B$-particles on site $i$
\begin{equation}
\frac{dB_{i}(t)}{dt}=J_{i}^{+}(t)-J_{i}^{-}(t)+\kappa A_{i}(t),
\label{sitebalance2}
\end{equation}
where $J_{i}^{+}$ denotes the gain flux and $J_{i}^{+}$ the loss flux
of particles $B$ at site $i$. The continuity equation reads:
\begin{equation}
\frac{dB_{i}(t)}{dt}=\frac{1}{2}J_{i-1}^{-}(t)+\frac{1}{2}
J_{i+1}^{-}(t)-J_{i}^{-}(t)+\kappa A_{i}(t).  
\label{balance2}
\end{equation}
A $B$-particle that leaves the site $i$ at time $t$ either has come there
as a $B$-particle at some prior time or was converted from an $A$-particle
that either was at site $i$ from the very beginning or arrived there later,
at $t^{\prime}>0$. Thus:
\begin{widetext}
\begin{eqnarray}
J_{i}^{-}(t)&&=\int_{0}^{t}\psi(t-t^{\prime })J_{i}^{+}(t^{\prime
})dt^{\prime}+\psi(t)\left[1-e^{-\kappa t}\right]A_{i}(0)+\int_{0}^{t}
\psi(t-t^{\prime})\left[1-e^{-\kappa(t-t^{\prime})}\right]
I_{i}^{+}(t^{\prime})dt^{\prime} \nonumber \\
&&=\int_{0}^{t}\psi(t-t^{\prime})\left[\frac{d B_{i}(t^{\prime})}
{d t^{\prime}}-\kappa A_{i}(t^{\prime})+J_{i}^{-}(t^{\prime})\right]
dt^{\prime}+\psi(t)\left[1-e^{-\kappa t}\right]A_{i}(0)+\nonumber \\
&&+\int_{0}^{t}\psi(t-t^{\prime})\left[1-e^{-\kappa(t-t^{\prime})}\right]
\left[\frac{dA_{i}(t^{\prime})}{d t^{\prime}}+\kappa
A_{i}(t^{\prime})+I_{i}^{-}(t^{\prime})\right]dt^{\prime},
\label{mainB}
\end{eqnarray}
where the local balance equation (\ref{sitebalance}) was
used. 
Applying Laplace transform and solving for $\tilde{J}_{i}^{-}(u)$ we
get:
\begin{equation}
\tilde{J}_{i}^{-}(u)=\frac{1}{1-\tilde{\psi}(u)}\left\{\tilde{\psi}(u)
\left[ u\tilde{B_{i}}(u)-\kappa
\tilde{A_{i}}(u)\right]+
\left[\tilde{\psi}(u)-\tilde{\psi}(u+\kappa)\right]\left[u\tilde{A_{i}}(u)
+\kappa\tilde{A_{i}}(u)+\tilde{I_{i}^{-}}(u)\right]\right\}.
\label{laplaceB}
\end{equation}
\end{widetext}
Here the initial condition $B_{i}(0)=0$ was explicitly used. 
Using Eq.(\ref{currentAlaplace}) for $\tilde{I}_{i}^{-}$ we get:
\begin{eqnarray}
\tilde{J}_{i}^{-}(u)=\tilde{\Phi}_{0}(u)\tilde{B}_{i}(u)+\left[\tilde{\Phi}_{0}(u)-
\tilde{\Phi}_{\kappa}(u)\right]\tilde{A}_{i}(u).
\end{eqnarray}
After inverse Laplace transformation we get:
\begin{eqnarray}
&& J_{i}^{-}(t) = \int_{0}^{t}\Phi_{0}(t-t^{\prime})B_{i}(t^{\prime})
dt^{\prime} + \nonumber \\
&& \: +\int_{0}^{t}\Phi_{0}(t-t^{\prime})\left[1-e^{-\kappa(t-t^{\prime})}
\right]A_{i}(t^{\prime})dt^{\prime}.
\end{eqnarray}
We now substitute this into the continuity equation (\ref{balance2}),
perform the transition to a continuum and get
\begin{eqnarray}
&&\frac{\partial  B(x,t)}{\partial t}
=\frac{a^{2}}{2}\int_{0}^{t}\Phi_{0}(t-t^{\prime})\Delta
B(x,t^{\prime})dt^{\prime}+\kappa A(x,t)+ \nonumber \\
&&\: +\frac{a^{2}}{2}\int_{0}^{t}\Phi_{0}(t-t^{\prime})
\left[1-e^{-\kappa(t-t^{\prime})}\right]\Delta A(x,t^{\prime})dt^{\prime}.
\label{Aequation}
\end{eqnarray} 
For the exponential waiting time density for which
$\Phi_{0}(t)=\lambda \delta(t)$ the third term in (\ref{Aequation})
vanishes and a normal reaction-diffusion equation arises.
For a power law waiting time distribution Eq.(\ref{Aequation}) can be
written in terms of fractional derivatives and the operator
$\hat{T}_{t}(1-\alpha,\kappa)$, Eq.(\ref{operatorT}):
\begin{eqnarray}
&&\frac{\partial B(x,t)}{\partial t}=K_{\alpha}\,_{0}D_{t}^{1-\alpha}
\Delta B(x,t)+\kappa A(x,t)+ \nonumber \\ 
&&\: +K_{\alpha} \left[ \,_{0}D_{t}^{1-\alpha}
- \hat{T}_{t}(1-\alpha,\kappa)\right] \Delta A(x,t).
\label{EqProd}
\end{eqnarray}
Note that the equation for the product contains the term depending on the
concentration of the component $A$ at all previous times.  
This term has to do with
the fact that the products are introduced into the system later on in
course of the reaction, and their motion therefore is described not by
the normal CTRW (and the corresponding fractional diffusion equation)
but by the aged one \cite{Aged}.  Note also that the sum of
Eqs.(\ref{Aconc}) and (\ref{EqProd}) always yelds the ``normal''
subdiffusion equation for the overall concentration $C(x,t)$,
Eq.(\ref{Sum}). Moreover the solutions of Eqs. (\ref{Aconc}) and 
(\ref{EqProd}) satisfy Eq. (\ref{E3}). 

In summary, we derived here the equations describing the time
evolution for the local $A$ and $B$-concentrations in a simple
monomolecular conversion reaction $A \rightarrow B$ taking place at a
constant rate $\kappa$ and under subdiffusion conditions. These 
equations do not have a usual form of
reaction-diffusion equations with the transport term independent on
the reaction one. This fact is due to nonmarkovian property of the
subdiffusion process, and will persist for more complex reaction
schemes as well.

IMS gratefully acknowledges the hospitality of the University of
Barcelona and the financial support by the HPC-Europa program.  Authors
are thankful to J. Klafter and R. Metzler for stimulating discussions.


\begin{thebibliography}{99}
\bibitem{generalbimol} For example, for an $A+B \rightarrow C$ reaction
which takes place locally at a rate $\kappa$ the survival probability $p$
of an $A$ particle in the time interval $[t^{\prime},t]$ is given by a
solution of the equation $dp/dt = -\kappa B(t)$. It reads:
$p(t)=\exp\left(-\int_{t^{\prime}}^{t}\kappa
B_{i}(t^{\prime})dt^{\prime}\right).$ The loss flux of $A$-particles
can then be written as follows:
$I_{i}^{-}(t)=\psi(t)\exp\left(-\int_{0}^{t}\kappa
B_{i}(t^{\prime})dt^{\prime}\right)A_{i}(0)+
\int_{0}^{t}\psi(t-t^{\prime})\exp\left(-\int_{t^{\prime}}^{t}\kappa
B_{i}(t^{\prime\prime})dt^{\prime\prime}\right)I_{i}^{+}(t^{\prime})dt^{\prime}.$
Note that this expression is a functional of the $B$-concentrations at
all previous times. The overall structure of equations in this case is
much more involved than in the monomolecular case considered in the
present Letter.
\end{thebibliography}
\end{document}